\documentclass[A4]{PoS}

\usepackage{subcaption}

\newcommand{\mnv}{MINER$\nu$A}

\title{Measurement of electron neutrino CCQE-like cross-section in MINERvA}

\ShortTitle{Measurement of electron neutrino CCQE-like cross-section in MINERvA}

\author{\speaker{Jeremy Wolcott}, for the MINERvA collaboration \\
        University Of Rochester, Rochester, New York 14610 USA\\
        E-mail: \email{jwolcott@pas.rochester.edu}}

\abstract{The electron-neutrino charged-current quasi-elastic (CCQE) cross-section on nuclei is an important input parameter to appearance-type neutrino oscillation experiments.  Current experiments typically work from the muon neutrino cross-section and apply corrections from theoretical arguments to obtain a prediction for the electron neutrino cross-section, but to date there has been no experimental verification of the estimates for this channel at an energy scale appropriate to such experiments.  We present a preliminary result from the MINERvA experiment on the first measurement of an exclusive reaction in few-GeV electron neutrino interactions, namely, the cross-section for a CCQE-like process.  The result is given both as differential cross-sections vs. the electron energy, electron angle, and $Q^{2}$, as well as a total cross-section vs. neutrino energy.}

\FullConference{16th International Workshop on Neutrino Factories and Future Neutrino Beam Facilities\\
		 25 -30 August, 2014\\
		 University of Glasgow, United Kingdom}

\begin{document}

	\section{Introduction}
		Current terrestrial neutrino oscillation experiments searching for fundamental information in the neutrino sector, such as the neutrino mass hierarchy and whether CP violation occurs for leptons, usually employ experimental designs which rely on the partial oscillation of a beam of muon neutrinos into electron neutrinos.\cite{T2K NIM,NOvA TDR}  These experiments build large detectors of heavy materials to maximize the rate of neutrino interactions, and then examine the energy distribution of the neutrinos that do interact with the detector, comparing the observed spectrum with predictions based on hypotheses of no oscillation or oscillation with given parameters.
		
		Correct prediction of the observed energy spectrum for electron neutrino interactions---on which these oscillation results depend---requires an accurate model of the rates and outgoing particle kinematics.  This, in essence, boils down to a need for precise $\nu_{e}$ cross-sections on the detector materials in use.  And yet, because of the difficulties associated with producing few-GeV electron neutrino beams, even when including very recent results, only two such cross-section measurements exist\cite{Gargamelle nue,T2K nue}.  Furthermore, the small statistics and inclusive nature of both of these measurements make their use as model discriminators challenging.  Instead, most simulations begin from the wealth of high-precision cross-section data available for muon neutrinos and apply corrections such as those discussed in ref. \cite{DayMcF} to obtain a prediction for $\nu_{e}$.
		
		We offer here a preliminary result in an effort to produce a higher-statistics cross-section for a quasi-elastic-like electron neutrino process, which is among the dominant reaction mechanisms at most energies of interest to oscillation experiments.  We use the \mnv{} detector, which consists of a central sampling scintillator region, built from strips of fluoror-doped scintillator glued into sheets, then stacked transverse to the beam axis; both barrel-style and downstream longitudinal electromagnetic and hadronic sampling calorimeters; and a collection of upstream passive targets of lead, iron, graphite, water, and liquid helium.  The detector design and performance are discussed in full detail elsewhere.\cite{MINERvA NIM}  \mnv{} occupies space in the NuMI $\nu_{\mu}$ beam, where it was exposed to a flux of $\sim 99$\% $\nu_{\mu}$ and $\sim 1$\% $\nu_{e}$ mostly between 3-5 GeV for this dataset.	We also compare the result for $\nu_{e}$ to a similar, previous \mnv{} result for $\nu_{\mu}$ to evaluate how similar they are.
	
	\section{Signal definition}
		In traditional charged-current quasi-elastic neutrino scattering, CCQE, the neutrino is converted to a charged lepton via exchange of a W boson with a nucleon, resulting in the following reaction: $\nu_{l} n \rightarrow l^{-} p$.  (Antineutrino scattering reverses the lepton number and isospin: $\bar{\nu}_{l} p \rightarrow l^{+} n$.)  Because the \mnv{} detector is not magnetized, we cannot differentiate between electrons and positrons on an event-by-event basis.  Moreover, hadrons exiting the nucleus after the interaction can re-interact and change identity or eject other hadrons\cite{GiBUU FSI}; furthermore, pairs of nucleons correlated within the initial state may cause multiple nucleons to be ejected by a single interaction\cite{Martini corr,Nieves corr}.  Therefore, we define the signal process ``phenomenologically,'' by its final-state particles: we search for events with either an electron or positron, no other leptons or photons, any number of nucleons, and no other hadrons.  We call this type of event ``CCQE-like.''  We also demand that events originate from a 5.57-ton volume fiducial volume in the central scintillator region of \mnv{}.
		
	\section{Event selection and backgrounds}
\begin{figure}[h]
			\centering
			\includegraphics[width=0.7\textwidth]{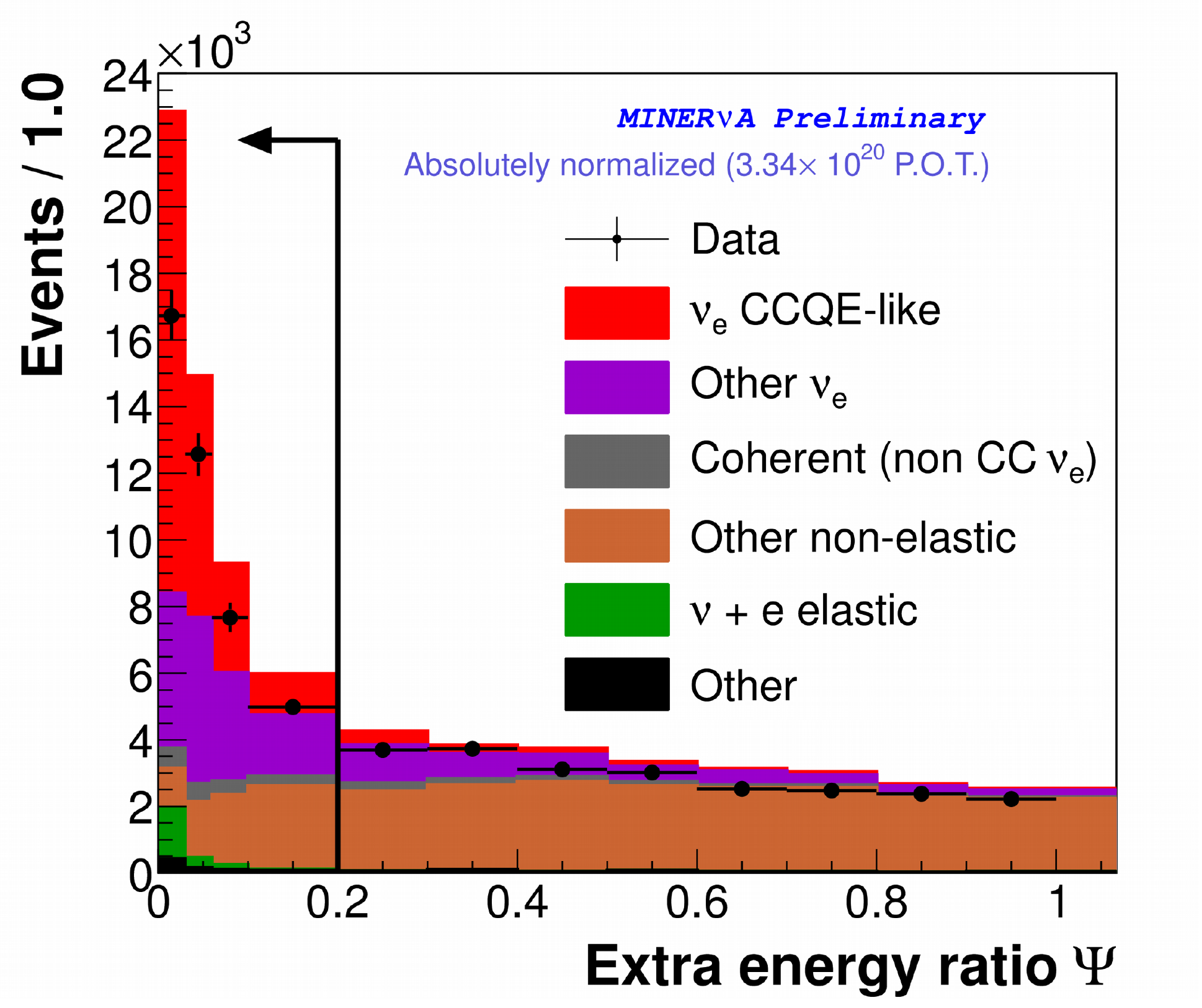}
			\caption{Sample cut on $\Psi$ (defined in the text) at the most probable event visible energy, $E_{vis} = 0.4$ GeV.}
			\label{fig:psi}
		\end{figure}

\begin{figure}[h]
			\centering
			\includegraphics[width=0.7\textwidth]{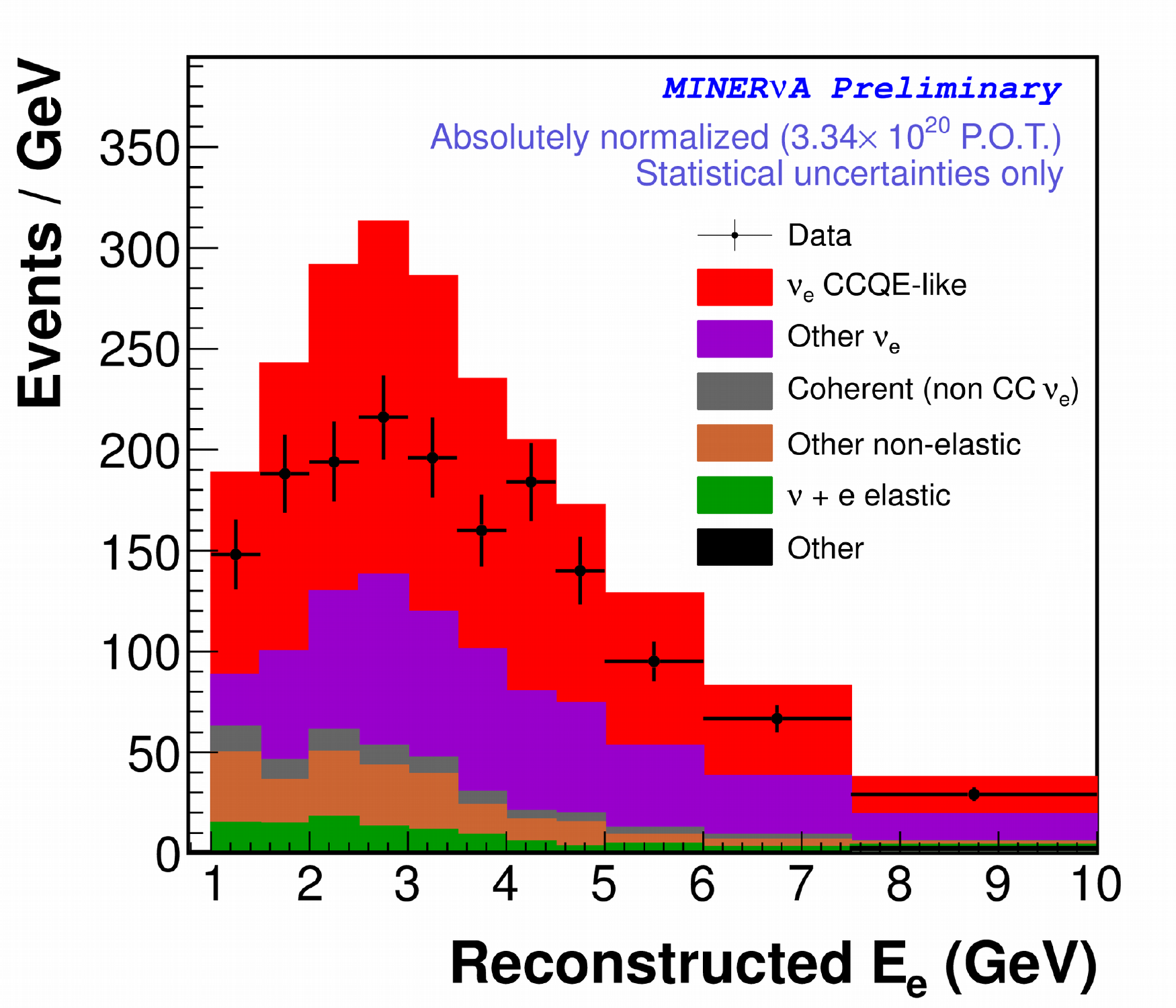}
			\caption{Event sample after all selection cuts.}
			\label{fig:selected sample}
		\end{figure}

		Candidate events are selected from the data based on three major criteria.  First, a candidate must contain a reconstructed cone object of angle $7.5^{\circ}$, originating in the fiducial volume, which is identified as a candidate electromagnetic cascade by a multivariate PID algorithm.  The latter combines details of the energy deposition pattern both longitudinally (mean $dE/dx$, fraction of energy at downstream end of cone) and transverse to the axis of the cone (mean shower width) using a $k$-nearest-neighbors (kNN) algorithm.  Secondly, we separate electrons and positrons from photons by cutting events in which the energy deposition at the upstream end of the cone is consistent with two particles rather than one (since photons typically interact in \mnv{} by producing an electron-positron pair).  At this point, the cone object becomes the electron candidate.  Our final criterion is an attempt to select CCQE-like interactions using a classifier we call ``extra energy fraction,'' $\Psi$, which, when an event's visible energy not inside the electron candidate or a sphere of radius 30cm centered around the cone vertex is denoted ``extra energy,'' is defined as:
		\begin{equation} \Psi = \frac{E_{\mathrm{extra}}}{E_{\mathrm{electron}}} \end{equation}
		Our cut is a function of the total visible energy of the event.  The cut at the most probable total visible energy, $E_{\mathrm{vis}} = 0.4$ GeV, is illustrated in fig. \ref{fig:psi}.  Finally, we retain only events with reconstructed electron energy in the range 1 GeV $\leq E_{e} \leq$ 10 GeV.  Here the lower bound excludes a region where the background estimate is still under further study, and the upper bound restricts the sample to events where the uncertainties on flux prediction are tolerable.  The distribution of events selected by this sequence is shown in fig. \ref{fig:selected sample}.
		
		As fig. \ref{fig:selected sample} shows, even after the final selection, a significant fraction of the sample is predicted to be from background processes.  We attempt to constrain the background model by examining sidebands in two of the variables already mentioned.  The first of these is in the $dE/dx$ measured at the front of the electron candidate; by choosing a sample at larger values, we can obtain a sideband rich in photon background events.  The second sideband is in the extra energy fraction $\Psi$; a sample of events at larger $\Psi$ constitutes a sideband rich in inelastic background.  We use these sidebands together to fit the normalizations of the three major backgrounds: $\nu_{e}$ non CCQE-like, non-$\nu_{e}$ coherent pion, and other inelastic events.  The three background classes' normalizations are fitted simultaneously, using distributions in both reconstructed candidate electron angle and energy, across the two sidebands, to obtain a single scale factor that represents the best estimate of the total normalization of the background as compared to the prediction from GENIE.  We obtain a scale factor of 0.69; this overall reduction is a similar trend to that observed when similar procedures were performed on other \mnv{} analyses.  Subsequent to the constraint, we scale the backgrounds in the signal region and subtract them from the data.  We then compare it to the simulated prediction of the signal process.

	\section{Cross-section result}
		We calculate three differential cross-sections in electron angle, electron energy, and four-momentum transfered from neutrino to nucleus $Q^{2}$, as well as the total cross-section vs. neutrino energy.  For neutrino energy and $Q^{2}$, we employ the commonly-used CCQE approximations (assuming a stationary target nucleon) which allow us to compute them from just the lepton kinematics:
		\begin{equation}
			E_{\nu}^{QE} = \frac{m_{n}^{2} - (m_{p} - E_{b})^{2} - m_{e}^{2} + 2(m_{p}-E_{b}E_{e})}{2(m_{p} - E_{b} - E_{e} + p_{e} \cos{\theta_{e}})}
		\end{equation}
		\begin{equation}
			Q^{2}_{QE} = 2 E_{\nu}^{QE} \left(E_{e} - p_{e} \cos{\theta_{e}}\right) - m_{e}^{2}
			\label{eq:q2}
		\end{equation}
		Differential cross-sections are calculated in bins $i$ according to the following rule for sample variable $\xi$, with $\epsilon$ representing signal acceptance, $\Phi$ the flux integrated over the energy range of the measurement, $T_{n}$ the number of targets (CH molecules) in the fiducial region, $\Delta_{i}$ the width of bin $i$, and $U_{ij}$ a matrix correcting for detector smearing in the variable of interest:
		\begin{equation}
			\label{eq:dsigma}
			\left( \frac{d\sigma}{d\xi} \right)_{i} = \frac{1}{\epsilon_{i} \Phi T_{n} \left(\Delta_{i}\right)} \times \sum_{j}{U_{ij} \left(N_{j}^{\mathrm{data}} - N_{j}^{\mathrm{bknd\ pred}}\right)}
		\end{equation}
		(The formula for the total cross-section differs only in that the flux is integrated only over the energy of bin $i$, rather than the whole energy range, and that we therefore do not need to divide by the bin width $\Delta_{i}$.)
		
		We perform unfolding in these four variables using a Bayesian technique\cite{D'Agostini unf} with a single iteration.  The unfolding matrices $U_{ij}$ needed as input are predicted by our simulation.  Our prediction for the neutrino flux $\Phi$ by which we then divide is derived from a GEANT4-based simulation of the NuMI beamline (described further in ref. \cite{antinumu PRL}).  In addition, we use an \textit{in situ} \mnv{} measurement based on elastic scattering of neutrinos from atomic electrons\cite{Jaewon JETP} to provide a data-based constraint for the flux estimate.  
		
\begin{figure}[h]
			\centering
			\includegraphics[width=0.7\textwidth]{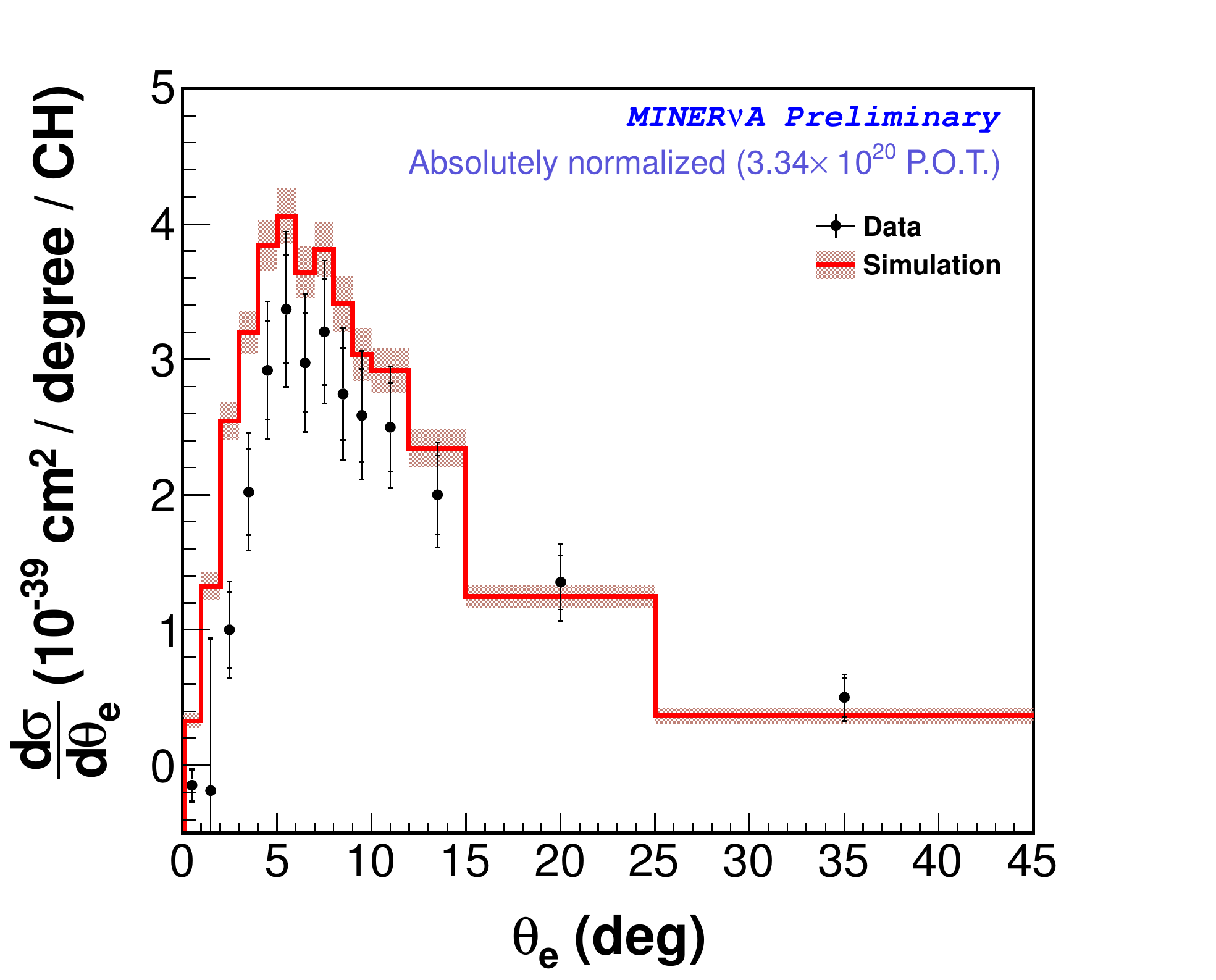}
			\caption{Differential cross-section vs. electron angle.  Inner errors are statistical; outer are statistical added in quadrature with systematic.}
			\label{fig:xs theta_e}
		\end{figure}

\begin{figure}[h]
			\centering
			\begin{subfigure}[c]{0.49\textwidth}
				\includegraphics[width=\textwidth]{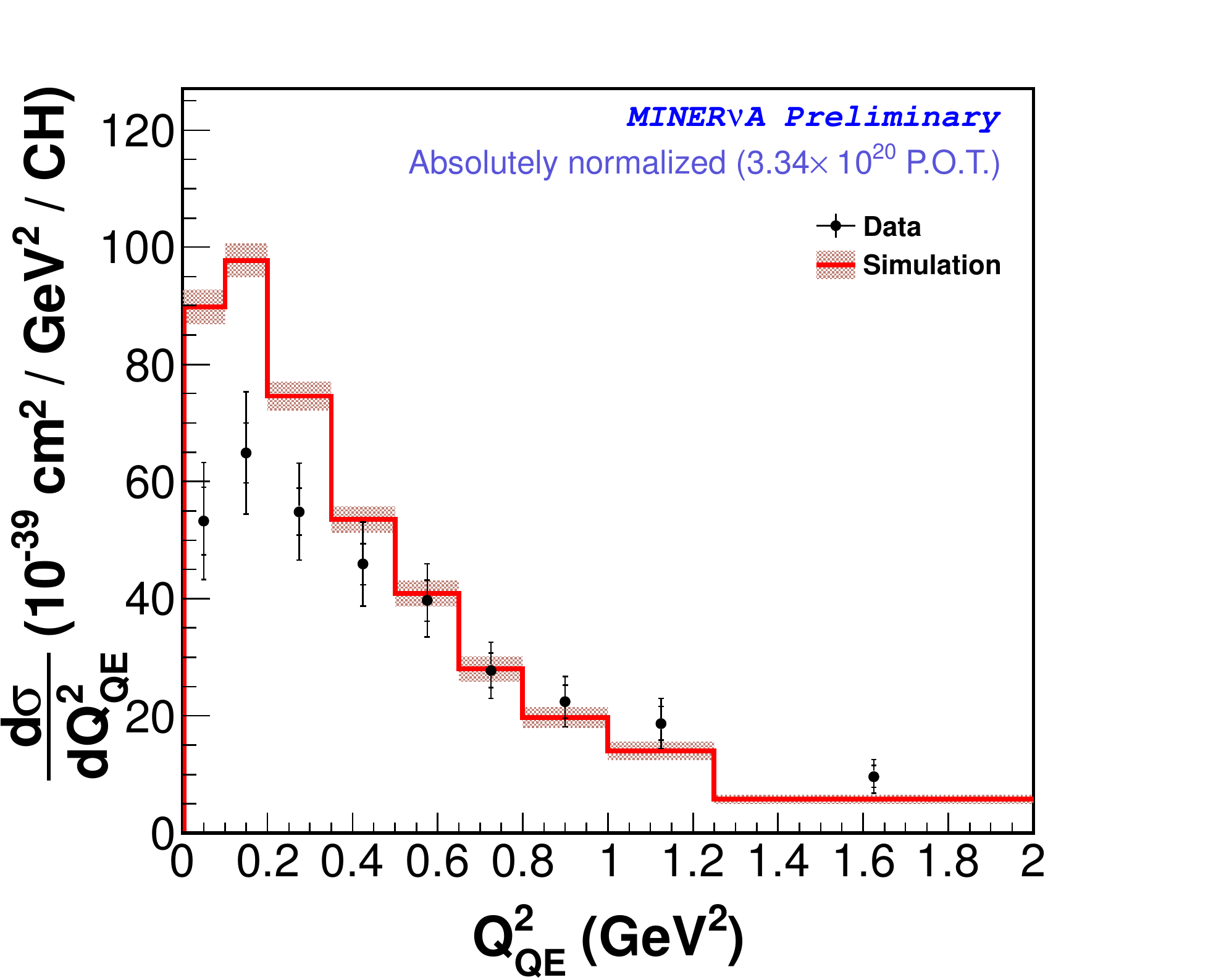}
				\caption{$\nu_{e}$}
				\label{fig:xs q2}
			\end{subfigure}
			\begin{subfigure}[c]{0.49\textwidth}
				\includegraphics[width=\textwidth]{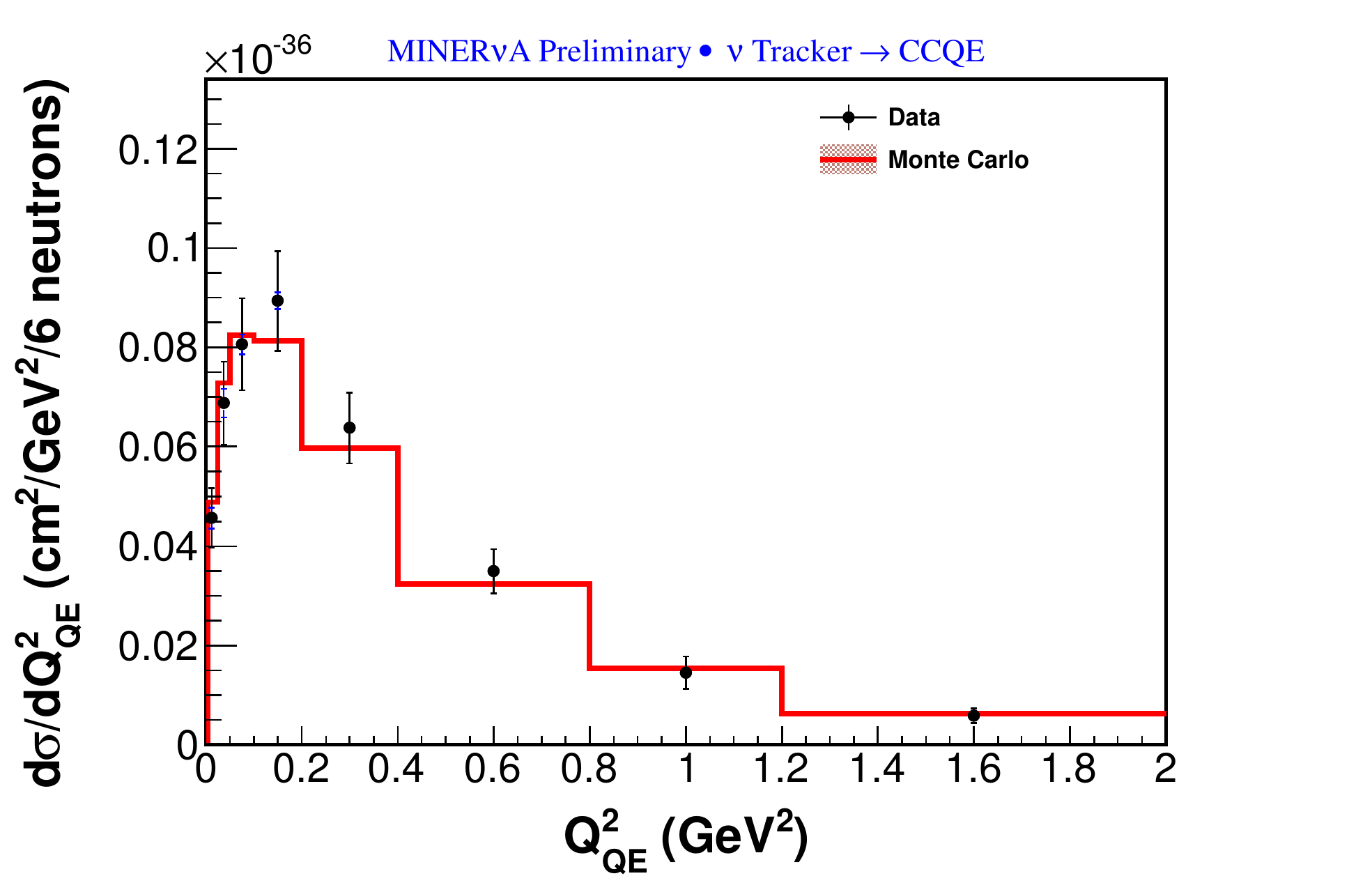}
				\caption{$\nu_{\mu}$ (measurement from ref. \cite{numu PRL})}
				\label{fig:xs q2 numu}
			\end{subfigure}

			\caption{Differential cross-section vs. $Q^{2}_{QE}$ (defined in the text) for electron and muon neutrinos.  Inner errors are statistical; outer are statistical added in quadrature with systematic.}
		\end{figure}

		The cross-section vs. electron angle obtained from this procedure is given in fig. \ref{fig:xs theta_e}.  We note that the simulation predicts substantially more events in the most forward bins than we observe in our data.  In addition, the measured cross-section vs. $Q^{2}_{QE}$, shown in fig. \ref{fig:xs q2}, appears to exhibit a noticeable migration from low to high $Q^{2}$ as compared to the prediction.  This is in contrast to the analogous cross-section measured in \mnv{} using muon neutrinos, fig. \ref{fig:xs q2 numu}, which agrees much better---if not perfectly---with GENIE's model.
		
	\section{Conclusions}
		Though the $\nu_{e}$ cross-section is vitally important for neutrino oscillation searches, experimental challenges have prevented extensive measurement of this quantity until recently.  In this preliminary result from \mnv{}, we observe a discrepancy at low angles between the model in GENIE 2.6.2 and our data in $d\sigma/d\theta_{e}$.  Furthermore, we find that the $Q^{2}_{QE}$ spectrum we observe appears to be harder for $\nu_{e}$ CCQE than it is for $\nu_{\mu}$ CCQE, in contrast to the prediction of GENIE.  If substantiated by further study, these observations will necessitate modifications to the models currently in use in neutrino generators so as to ensure they correctly simulate the electron neutrino kinematics.  Work is still ongoing to characterize the backgrounds in the $E_{e} < 1$ GeV region, and a full result will be published soon.

\end{document}